\documentclass[10pt]{article}
\usepackage{graphicx}
\usepackage[english]{babel}
\usepackage{amsfonts,latexsym}
\usepackage{amsmath}
\usepackage{epstopdf} 
\usepackage{latexsym}
\usepackage[latin1]{inputenc}

\def\ca{\cite{ashby18a}}
\def\be{\begin{equation}}
\def\ee{\end{equation}}
\def\ba{\begin{eqnarray}}
\def\ea{\end{eqnarray}}

\begin{document}
\title{
Reply to comment by S. Svitlov \cite{svitlov18} \ on \emph{Relativistic theory of the falling cube gravimeter}$^*$}

\author{Neil Ashby$^{1,2}$ \\ $^1$National Institute of Standards and Technology,\\
Boulder,CO 80305 USA\\
$^2$University of Colorado, Boulder, CO USA\\
E-mail:  ashby@boulder.nist.gov}
\maketitle
\begin{abstract}

	In the subject paper \cite{ashby18} of the comment \cite{svitlov18}, light propagation through an absolute gravimeter was analyzed, including the propagation delay through the falling retroreflector and through the vacuum. The resulting expression for the interference signal applies without any subsequent ``speed-of-light" correction.  Other corrections appeared for the three fitting parameters $Z_0,\ V_0$ and $g$, which are the initial position and velocity, and the acceleration of gravity at the reference point.  The comment assumes the value of $Z_0$ is known apriori; this case was not addressed in \cite{ashby18}.    Also, the comment misunderstands statements made in \cite{ashby18} regarding the derivation of the relativistic/nonrelativistic parts of the corrections (Eq. (47) of \cite{ashby18}), and mistakenly claims they apply to the undifferenced interference signal.  In this reply we show why, because of inapplicable assumptions, approximations and misunderstandings, the comment does not apply to the results of \cite{ashby18}\,.
\end{abstract}

\vspace{2pc}

\renewcommand{\thefootnote}{\fnsymbol{footnote}}
\setcounter{footnote}{1}
\footnotetext{Work of the U.S. government, not subject to copyright.}
\setcounter{footnote}{0}
\vspace{2pc}

The purpose of \cite{ashby18} was to provide a relativistic analysis of the propagation of light through a falling retroreflector gravimeter, including propagation through the retroreflector.  The result was an expression for the interference signal at the recombination point (Eq. (33) of \cite{ashby18} with the reference beam phase removed). Accounting for propagation delay in the solid retroreflector results in several contributions to the final interference signal involving properties of the retroreflector: the corner to face distance $D$, center of mass-to-face distance $d$, and refractive index $n$. The interference signal is a fifth degree polynomial in the time $T$.  The advantage of the relativistic expression is that no ``speed-of-light" correction is needed, as all propagation delays are accounted for.  

In \cite{ashby18} data from 5000 drops was first fit with the fully relativistic expression, then fit using the relativistic expression with $D$ and $d$ set to zero, and differences in the resulting values of $g$ and the other fitting parameters were reported, which turned out to be the same in every drop.  A theoretical explanation for these differences was obtained by taking corresponding differences of the theoretical interference signal; agreement between the numerical differences and the theoretical differences was excellent.  The comment \cite{svitlov18}\ contradicts the conclusions reported in \cite{ashby18}, primarily because the comment assumes that the initial retroreflector position parameter $Z_0$ is known a priori, but also because statements made in \cite{ashby18} have been misinterpreted.  Fitting a model to data is independent of the derivation of the model; choosing which parameters to include in the fit to the model may or may not lead to a good fit.  In fact in the data processed and reported on in \cite{ashby18} no a priori value for $Z_0$ was available; the fit value for $g$ would depend critically on the value of $Z_0$ inserted into the model.
	
It is worthwhile to summarize the theory leading to the differences derived in \cite{ashby18}.  A polynomial model of degree $N$ with time $T$ as independent variable, which is fit to a given data set, will return specific numerical values of the polynomial coefficients.  If there are two different models of the same degree for the data, the coefficients may have different interpretations but will be numerically equal.  The numbers then have to be interpreted in terms of the model. In the present case we are comparing two models of the same degree based on the interference signal derived in \cite{ashby18}.  To compress the algebra, we work with the interference signal multiplied by $c/(2 \Omega)$ where $c$ is the speed of light and $\Omega$ is the angular frequency of the reference beam light.  We write this function as
\be
F(Z_0,V_0,g,T,L)=\frac{c}{2\Omega}\phi(Z_0,V_0,g,T,L)\,.
\ee
where $\phi(Z_0,V_0,g,T,L)$ is the interference signal and $L=Dn-d$.  In the model adopted in \cite{ashby18}\ the center of mass (COM) motion as a function of time $T$ is
\be\label{com}
Z_{cm}(T)=Z_0+V_0 T-\frac{1}{2} g T^2 +\gamma\big(\frac{Z_0 T^2}{2}+\frac{V_0 T^3}{6}-\frac{g T^4}{24}\big)\,,
\ee 

  It was shown in \ca\  that $Z_{nr}=Z_{cm}+L$ represents a fictitious point above the corner from which the actual interference signal can be inferred by considering the time required for the signal from the fictitious point to pass through a vacuum with speed $c$ to the point of recombination.  The quantity $Z_{nr}(T)$ simply contains all the ``non-relativistic" contributions that were derived in Appendix B of \cite{ashby18}.  After computing $Z_{nr}$ at the retarded time, expanding in powers of $c^{-1}$ while keeping the first term in $c^{-1}$, the interference signal at the recombination point can be compactly expressed in terms of $F$ with
\be\label{signal}
F(Z_0,V_0,g,T,L)=(Z_{cm}+L)\bigg(1-\frac{V_{cm}}{c}  \bigg)\,,
\ee
where $V_{cm}$ is the COM velocity obtained by differentiating Eq. (\ref{com}).  Terms of degree higher than unity in $\gamma$ are neglected\,.  The ``relativistic" part of the signal can be identified from Eq. (\ref{signal}) as the part containing $c^{-1}$.

For brevity we define a difference operator $\Delta$ acting on a polynomial by
\ba\label{diff0}
\Delta F(Z_0,V_0,g,T,L) \equiv \hbox to 2.0 truein{} \notag\\
F(Z_0+\delta Z_0,V_0+\delta V_0,g+\delta g,T,L)-F(Z_0,V_0,g,T,0)=0\,.
\ea
The two models compared are: the first term, which includes optical properties of the prism, and the second term, which omits them.  In neither case is any additional ``speed of light correction" relevant.  Both models are derived from, but are not equivalent to, the assumed center of mass (COM) motion of the prism, Eq. ({\ref{com}})\,. The quantities $\delta Z_0,\ \delta V_0, \ \delta  g$ represent small differences for the three free parameters of interest. If they are sufficiently small that Eq. (\ref{diff0}) can be linearized, this becomes
\be\label{diff1}
\Delta F(Z_0,V_0,g,T,L) = \frac{\partial F}{\partial Z_0}\delta Z_0+
\frac{\partial F}{\partial V_0}\delta V_0+
\frac{\partial F}{\partial g}\delta g + \frac{\partial F}{\partial L }L=0\,,
\ee
the last term arising because the signal is linear in $L$.  The difference is zero numerically from fitting to the same data so the net coefficient of each power of $T$ in Eq. (\ref{diff1}) must vanish.  Because both models are polynomials of fifth degree in $T$, six conditions on the corrections are obtained. 


We give two examples: the coefficients of $T^0$ and $T^4$. First for the sum of all constant terms in Eq. (33) of \cite{ashby18} we have
\be
\Delta\bigg((Z_0+L)\big(1-\frac{V_0}{c}\big)\bigg)\,=
\big(1-\frac{V_0}{c}\big)\delta Z_0-\frac{(Z_0+L)}{c}\delta V_0+\big(1-\frac{V_0}{c}\big)L=0\,.
\ee
The coefficient of $T^4$ in the interference signal $F$ gives
\be\label{deltaT4term} 
\Delta\bigg(\gamma\bigg(-\frac{g}{24}+\frac{5 g V_0}{8c}\bigg)\bigg) =
\frac{\gamma}{24}\bigg(-1+15 \frac{V_0}{c}\bigg)\delta g+\gamma\frac{5 g}{8 c}\delta V_0=0\,.
\ee
When $\delta g$ is proportional to $\gamma$, both terms in Eq. (\ref{deltaT4term}) will be negligible.  Proceeding in this way through all six coefficients, the equations can be written compactly by introducing a matrix $A$ representing the non-relativistic part of the expressions, and $B/c$ representing relativistic contributions.  These matrices are defined as follows:
\be\label{nonrelativistic}
A=\left(
\begin{matrix}
1&0&0\\
0&1&0\\
\frac{\gamma}{2}&0&-\frac{1}{2}\\
0&\frac{\gamma}{6}&0\\
0&0&-\frac{\gamma}{24}\\
0&0&0
\end{matrix}
\right)
\ee
\be\label{relativistic}
B=\left(
\begin{matrix}
-V_0&-(L+Z_0)&0\\
g-\gamma(L+2Z_0)&-2V_0&L+Z_0\\
\mathstrut-2\gamma V_0& \frac{3g}{2}-\frac{\gamma L}{2}-2 \gamma Z_0&\frac{3V_0}{2}\\
\mathstrut\frac{7 \gamma g}{6}&- \frac{4\gamma V_0}{3}& -g +\frac{\gamma(L+7Z_0)}{6}\\
\mathstrut0& \frac{5  \gamma g}{8}&\frac{5 \gamma V_0}{8}\\
\mathstrut 0&0&-\frac{ \gamma g}{4}
\end{matrix}
\right)
\ee
The system of equations that express equality of the coefficients of each power of $T$ are then
\be\label{sixequations}
\bigg(A+\frac{1}{c}B\bigg)
\left(
\begin{matrix}
\delta Z_0\\
\delta V_0\\
\delta g
\end{matrix}
\right)
-\left(
\begin{matrix}
\mathstrut-L+\frac{L V_0}{c}\\
\mathstrut-\frac{gL-\gamma L Z_0}{c}\\
\mathstrut\frac{\gamma L V_0}{2c}\\
\mathstrut-\frac{g\gamma L}{6 c}\\
0\\
0
\end{matrix}\right)=0\,.
\ee

 The rank of the matrix $A+B/c$ is 3 (as are the ranks of $A$ and $B$ separately), so three independent equations can be selected, say the first three, corresponding to powers of $T^0,T,T^2$\,. The solutions that result are 
\ba\label{finalsolutions}
\delta Z_0=-L;\notag\\
\delta V_0=0;\\
\delta g=\gamma \delta Z_0=-\gamma L \notag\,,
\ea
and these are the differences to which the fitting routines converge.  Substituting these solutions into the remaining three equations, corresponding to powers $T^3,\ T^4,\ T^5$, gives respective residuals proportional to the small quantity $\gamma^2$:  
\be\label{residuals} 
\big\{0,0,0,-\gamma^2\frac{L(L+7 Z_0)T^3}{6c},\gamma^ 2L\frac{\big(c-15V_0\big)}{24c}T^4,\gamma^2\frac{g LT^5}{4 c}\big\}\,.
\ee
Such higher degree terms in $\gamma$ have been neglected throughout \cite{ashby18}; numerical values of the phase residuals for $T\approx 0.3$ s are less than $10^{-8}$ radians and are negligible. The derived differences, Eqs. (\ref{finalsolutions}), agree with differences obtained from numerical fits, showing in two ways that the optical properties of the retroreflector are significant, not negligible as claimed in the comment \cite{svitlov18}.  No substantive critique of the interference phase signal derivation, or of the steps described above that lead to the differencess, Eqs. (\ref{finalsolutions}), have been encountered yet.  We now discuss a selected set of issues we have with the comment \cite{svitlov18}.  

	The comment assumes $Z_0$ is known a priori, or $\delta Z_0=0$.  It is then logically inconsistent to apply the result for $\delta g$ from Eq. (\ref{finalsolutions}), as in Eq. (20) of \cite{svitlov18}, because the results in Eq. (\ref{finalsolutions}) are related to each other by the linearization process described above. A different analysis is required, which was not given in \cite{ashby18}.
	 
 Terms in Eq. (6) of the comment \cite{svitlov18}\  that are proportional to $T, T^2,$ and  $T^3$ were neglected there. But they are important for reducing residuals to negligible levels, Eqs. (\ref{residuals}).  They are not negligible. 

 Two facts were discovered while seeking an explanation for the numerical value of $\delta g$.  First, keeping only the relativistic terms in Eq. ({\ref{sixequations}})--those having the factor $c^{-1}$--the same solutions Eqs. (\ref{finalsolutions}) result.  Second, if all relativistic effects were omitted by letting $c \rightarrow \infty$, the difference operation Eq. (\ref{diff1}) or ({\ref{sixequations}) leads again to Eqs. (\ref{finalsolutions}). The comment \cite{svitlov18}\ misunderstands these results, asserting that it confirms ``the relativistic time-dependent dynamical effect due to the light propagation within the cube is negligible."  The misunderstanding mistakenly elevates the vanishing of a difference between relativistic terms to the vanishing of all the relativistic terms in the interference signal.  In reporting numerical differences in \cite{ashby18}, relativistic terms were not neglected.
  
In Eqs. (2-6) of \cite{svitlov18} the so-called ``displacement" (the signal, Eq. (\ref{signal}),  is not a displacement) is divided into the sum of four contributions.  Such divisions are arbitrary since any function of the time could be added to one of the contributions as long as it is subtracted from another.  For example, it is of particular interest to write all of the non-relativistic contributions (which were derived separately in Appendix B of \cite{ashby18}) in one component of the model, such as
\be\label{Znonrel}
Z_{nr}(Z_0,V_0,g,T,L)=Z_{cm}(Z_0,V_0,g,T)+L\,,
\ee
instead of grouping the term $L$ elsewhere and using only $Z_{cm}(T)$ as in the comment.   
For example, Eq. (\ref{Znonrel}) by itself when linearized and the difference with $Z_{cm}(Z_0,V_0,g,T)$ is taken gives:
\be\label{linearZnonrel}
\Delta\big(Z_{nr}(Z_0,V_0,g, T,L)\big)=0\,,
\ee
and leads to the solutions, Eqs. (\ref{finalsolutions}). However the model displacement of the COM motion used in \cite{ashby18} was actually $Z_{cm}(T)$, not $Z_{nr}(T)$ as claimed in \cite{svitlov18}. 

	The first ``displacement" component considered by \cite{svitlov18} (Eq. (2) of \cite{svitlov18}), is just the COM  position, $Z_{cm}(T)$, of the falling prism.  The comment asserts that this expression ``delivers the unbiased estimate of $g$."  But the comment assumes the value of $Z_0$ is known a priori, (see Sect. 4 in \cite{svitlov18}, first paragraph) ``being measured for every setup of an individual absolute gravimeter, either with the glass prism or with a hollow retroreflector."  The constraint $\delta Z_0=0$ was not considered in \cite{ashby18}; it creates a separate problem deviating in many ways from \cite{ashby18}.  If $Z_0$ were known a priori to within a few microns the fitting routines would be different and it is not clear that linearization is appropriate.  Results from two such different situations lead to errors if combined as in Eq. (20) of \cite{svitlov18}, which contradicts Eq. (\ref{linearZnonrel}) above.  On the other hand user manuals for such gravimeters give the equivalent of the COM position $Z_{cm}(T)$ for the model of free fall \cite{lacoste14}, the constants being ``free parameters providing the best estimates for initial position, velocity and gravity."  In many references $Z_0$ is regarded as an adjustable parameter \cite{nagornyi11,niebauer95,rothleitner10}.  Measurements can  provide a good estimate for $Z_0$ but allowance is usually made for variations in the conditions of each individual drop.  In \cite{ashby18}, fitting determined $Z_0$ with uncertainties of a few microns; it seems unlikely that a priori measurements could attain such accuracy; the COM is inside the glass.  
	
	   The coefficient of $T^2$ in Eq. (\ref{Znonrel}) is proportional to $g-\gamma Z_0$; this is the prism acceleration at $T=0$.  If there is no correction to the unbiased value of $Z_0$ in the fits there will be no correction to $g$. The comment claims that the measured value of $g$ relates to $Z=L$ rather than to Z=0.  This is incorrect as the computation consistently fits the model to the given data, with $g$ denoting its value at the reference position $Z=0$. The quantity $\delta Z_0$ refers to a correction for the initial position at $T=0$.  
	   
	   In Eq. (20) of the comment an expression is given claiming to represent the actual fitted model for the 5000 drops reported in \cite{ashby18}. The expression is incorrect in more than one way.  The data were fitted to a model based on $Z_{cm}$, but given in Eq. (\ref{signal}) above or Eq. (33) of \cite{ashby18}, including relativistic effects.  In Eq. (20) of \cite{svitlov18} the value of $g$ has been adjusted but it was assumed that $\delta Z_0=0$, so results reported in \cite{ashby18}\ can't be reliably used in that equation.  

At the end of Sect. 2 of \cite{svitlov18}\ it is claimed that the time-dependent dynamical effect due to the light propagation within the cube is negligible, confirmed by the fact that the ``same number (-6.8 $\mu$Gal) arises if all terms of order $c^{-2}$ are neglected."  Such terms were actually not neglected in the fits; the quote from \cite{ashby18}\  only describes results of taking differences of the nonrelativistic part of the signal.  As shown above, it is possible to pull the expression apart into relativistic and non-relativistic parts and discuss individual parts but the full expression (with and without $L$) was actually used for fitting. It is a misunderstanding of the attempt to understand differences, Eq. (\ref{sixequations}), by looking at parts of the difference separately.  Considering them together is sufficient.

An interference phase with all relativistic terms omitted is the subject of Sect. 3 of the comment.  This section is based on a misunderstanding of the statement quoted in the preceding paragraph, which was intended to express the fact (derived and discussed above), that if all ``relativistic" terms in Eq. (\ref{sixequations}) were omitted, the difference solutions Eqs. (\ref{finalsolutions}) would still result.  The use of an interference phase as in Eq. (12) of the comment is not relevant as such a phase was not used in the fitting reported in \cite{ashby18}.  

In Eq. (21) of the comment an expression for the interference signal ``as it is seen by the photodetector and is used for counting fringes" is provided in which the argument of the trigonometric functions neglects all relativistic effects.  The phase used in \cite{ashby18} had higher degree polynomial terms, up to $T^5$ (up to $T^3$ multiplying $L$), so conclusions following from Eq. (21) of the comment don't apply to results reported in \cite{ashby18}.

In sum, \cite{ashby18} accounts for light propagation in the falling retroreflector and predicts small differences for the fitted values for $g$ and $Z_0$, when properties of the prism are included or omitted respectively, but with no speed-of-light correction in either case. The interference signal derived in \cite{ashby18}\ can support various observables.   For example, a frequency observable could be obtained by differentiating the signal with respect to time.  Observables could be constructed from first or second differences in the time or for three-level schema, etc. No additional speed-of-light correction is needed.  The comment \cite{svitlov18} accepts the interference signal derived in \cite{ashby18}, then rearranges terms, makes certain assumptions and approximations, and revives the unnecessary speed of light correction. Misunderstandings of \cite{ashby18} are a serious problem within \cite{svitlov18}.   Because the comment assumes $Z_0$ is known a priori and misunderstands statements about differences resulting from comparison with and without the optical quantity $Dn-d$, the comment does not provide a valid critique of \cite{ashby18}.

\end{document}